\documentclass[prl,twocolumn,showkeys,showpacs,superscriptaddress]{revtex4}
\usepackage[OT1]{fontenc}
\usepackage[latin1]{inputenc}
\usepackage{color}
\usepackage{graphicx}
\usepackage{amssymb,amsmath}
\usepackage{upgreek}
\newcommand{\Ein}{E_{\mathrm{in}}}
\newcommand{\Eout}{E_{\mathrm{out}}}

\newcommand{\Ioutoff}{I_{\mathrm{out}}^{\mathrm{off}}}

\newcommand{\Eouton}{E_{\mathrm{out}}^{\mathrm{on}}}

\newcommand{\Eoutoff}{E_{\mathrm{out}}^{\mathrm{off}}}

\newcommand{\vbar}{\overline{v}}
\newcommand{\erfc}{\mathrm{Erfc}}
\newcommand{\bEout}{\overline{E_{\mathrm{out}}}}
\newcommand{\tEin}{\tilde{E}_{\mathrm{in}}}
\newcommand{\tEout}{\tilde{E}_{\mathrm{out}}}
\newcommand{\tEinon}{\tilde{E}_{\mathrm{in}}^{\mathrm{on}}}
\newcommand{\tEinoff}{\tilde{E}_{\mathrm{in}}^{\mathrm{off}}}

\begin{document}

\title{Coherent flash of light emitted by a cold atomic cloud}
\author{M. Chalony}
\affiliation{Institut Non Lin\'eaire de Nice, Universit\'e de Nice Sophia-Antipolis, CNRS, 06560 Valbonne, France}
\author{R. Pierrat}
\affiliation{Institut Langevin, ESPCI ParisTech, CNRS UMR 7587, 10 rue Vauquelin, 75005 Paris, France}
\author{D. Delande}
\affiliation{Laboratoire Kastler Brossel, UPMC-Paris 6, ENS, CNRS; 4 Place Jussieu,
75005 Paris, France}
\author{D. Wilkowski}
\affiliation{Institut Non Lin\'eaire de Nice, Universit\'e de Nice Sophia-Antipolis, CNRS, 06560 Valbonne, France}
\affiliation{Centre for Quantum Technologies, National University of Singapore, 117543 Singapore, Singapore}

\date{\today{}}

\begin{abstract}
When a resonant laser sent on an optically thick cold atomic cloud is
abruptly switched off, a coherent flash of light is emitted in the forward direction. This transient
phenomenon is observed due to the highly resonant character of the atomic scatterers. We analyze quantitatively
its spatio-temporal properties and show very good agreement with theoretical
predictions. Based on complementary experiments, the phase of the coherent field is reconstructed without interferometric tools.
\end{abstract}



\maketitle

\begin{figure}
\begin{center}
\includegraphics[scale=0.75]{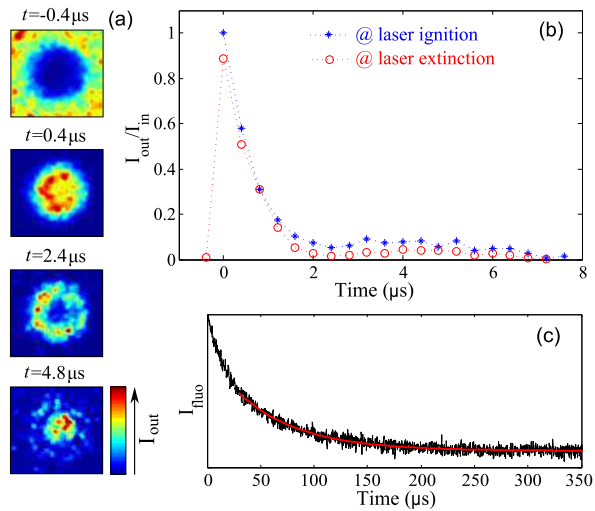}
\caption{(color online)
(a): Images of the signal transmitted by a cold cloud with maximum optical thickness (at the center) $b=6.5$ and temperature $T=3.8\,\upmu\textrm{K}$.
The images are taken downstream during the laser extinction. The first image ($t=-0.4\,\upmu$s)
shows, in the stationary regime, the shadow of the atomic cloud which has scattered most
of the incoming beam.
After the incoming laser is switched off at time $t=0$, a coherent flash of light -- with a spatio-temporal structure -- is emitted
by the atomic cloud, decaying over few $\upmu\textrm{s}$.
(b): Coherent transmission at the laser ignition and extinction at the cloud center. (c): Temporal evolution of the incoherent fluorescence intensity (black line) at the laser extinction. The red curve is a fit by a decreasing exponential.}
\label{Time_Evo}
\end{center}
\end{figure}

When sent on a turbid object, a coherent light beam is scattered by the heteoregeneities of the object.
If the mean free path $\ell$ of light inside the scattering object is shorter than its thickness $L,$ one enters into the multiple scattering regime where the photons follow a random walk. Nevertheless, a small part of the incoming photons cross the medium ballistically producing a coherent transmission equal to $e^{-b},$ where $b=L/\ell$ is the optical thickness of the medium. When $b$ is large, it seems thus hopeless to coherently transmit a laser beam through an optically thick medium. In this letter, we show experimentally that this is not true and that coherent transmission almost equal to 100\% can be achieved, although only over a short temporal window [see e.g. fig.~\ref{Time_Evo}b].

At the microscopic level, the coherently transmitted field can be seen as the result of the destructive interference between the incoming laser beam and the field radiated  in the forward direction by the dipoles induced in the medium. In the stationary regime, the depletion of the coherent beam is exactly compensated by the intensity scattered in other directions. The simple idea of our experiment is to abruptly switch off a monochromatic laser beam incoming on an optically thick medium. If the laser extinction is fast enough, the induced dipoles continue to radiate a coherent field, creating a coherent flash of light decaying over their lifetime.

The extremely fast response time (fs or ps) of standard dielectric scatterers make the observation of the coherent flash challenging. This issue can be in principle overcome using resonant scatterers with long dwell time. For this reason, the first reported observation of coherent emission after the source extinction -- known as free induction decay (FID)-- came from NMR few decades ago \cite{hahn1950nuclear}. In the optical domain, observation of FID in an optically thin medium was reported on a molecular thermal gas \cite{brewer1972optical,foster1974interference}. More recently, optical FID in a cold cloud of rubidium gas, without inhomogeneous broadening, was also observed and compared with Maxwell-Bloch equations in the zero temperature limit \cite{toyoda1997optical,Shim2002,Jeong2006,Wei2009}. In this letter, FID in an optically thick medium is studied, with emphasis on the effect of the optical thickness and temperature.

A cold atomic sample of Strontium with up to $10^7$ atoms in a volume of about $0.01\,\textrm{mm}^3$, is produced in a magneto-optical trap
on the $^1\!S_0\rightarrow\!^3\!P_1$ intercombination line at $\lambda=689\,\textrm{nm}$ with a excited lifetime of $\Gamma^{-1}=21\upmu\textrm{s}$~\cite{chaneliere2008three}.
The minimal temperature of the sample is $0.7\,\upmu\textrm{K}$ which corresponds to $k\vbar\simeq 1.6\Gamma$, where $\vbar$ is
the mean atomic velocity (along any direction). Thus, in contrast to standard laser cooling,
Doppler effect still broadens the atomic transition.
After loading, the magneto-optical trap is switched off.
A bias B-field of $1\,\textrm{G}$ is applied and a probe laser at exact resonance with the $^1\!S_0\rightarrow\!^3\!P_1,m=0$ is turned on.
The waist of the probe laser is $0.5\,\textrm{mm}$, so it can be well approximated by a plane wave. The probe is kept on during $40\,\upmu\textrm{s}$
at an intensity lower than half the saturation intensity of the transition ($3\,\upmu\textrm{W/cm}^2$).
On average, each atom scatters less than $0.3$ probe photons, ensuring that mechanical effects are very small. The ignition and the extinction of the probe laser are achieved in less that $50\,\textrm{ns}$.  The forward transmitted light is collected on an intensified CCD, placed on an image plane of the cloud, with a spatial resolution of $20\,\upmu\textrm{m}$. The gate time is $400\,\textrm{ns}$. The imaging system is well adapted for spatio-temporal studies. Moreover, the collection solid angle, $5\times 10^{-3},$ is small enough for the incoherent fluorescence detected by the camera to be negligible.

Some keys results are presented in fig.~\ref{Time_Evo}.
In (a), the images show the shadow of the atomic cloud in the stationary regime (upper image), due to the depletion of the
incoming laser beam by the atomic scatterers, and the flash of light emitted by the medium after the laser
extinction (lower images). Note the ring structure, discussed below.
Fig.~\ref{Time_Evo}(b)  shows the temporal evolution of the coherent transmission at the cloud center after laser ignition and extinction. The intensity of the coherent flash immediately after extinction is almost equal to the incoming laser intensity while the coherent transmission in the stationary regime (just before switch off) was negligibly small.
Fig.~\ref{Time_Evo}(c) shows the incoherent fluorescence signal, recorded at a angle ($30^{\circ}$) from the forward direction -- so that the coherent flash does not contribute --
after the laser extinction.
The decay is exponential with a characteristic time $\tau_{\textrm{fluo}}=57\,\upmu\textrm{s}=2.7\Gamma^{-1}$.
As a single scattering event is characterized by a dwell time $\Gamma^{-1},$
this proves that multiple scattering and radiation trapping are at play \cite{fioretti1998observation,labeyrie2003slow}.
The rather small value $\tau_{\textrm{fluo}}$ suggests that multiple scattering remains limited.
This is due to the ``large'' Doppler effect that can bring the scattered photon out of resonance, so that it can escape the medium before
being rescattered. We performed Monte Carlo simulations~\cite{pierrat2009enhancement}
which include both Doppler and recoil effects which turn out to be in excellent agreement with the observed decay time.

In sharp contrast, typical transient times for the coherent flash, see fig.~\ref{Time_Evo}(b), are \emph{much} shorter.
It is a clear signature that it has a different physical origin than radiation trapping, and that dephasing phenomena occur at a time scale shorter than $\tau_{\textrm{fluo}}$. In the sequel of this paper,
we analyze the physical processes responsible for the loss of phase coherence and show that they explain
quantitatively the experimental observations.
A first process is the finite lifetime of the atomic excited state related to the finite width of the atomic resonance. A second effect is due to the residual atomic motion.
Once the laser is turned off, the atomic motion leads
to dephasing of the dipoles, with a time constant expected to be proportional to $(k\vbar)^{-1},$ the time needed for an atom to travel one laser wavelength.
A third identified dephasing mechanism occurs in a thick medium: deep inside the medium, the atomic dipoles are driven by the superposition of the incident field and the field radiated by other atoms. Thus, the phase coherence
between the various layers is more fragile, producing a shorter coherent flash.

We now sketch the principle of the theoretical calculation. For the sake of simplicity, we assume that
the laser beam is a plane wave with wavevector $\vec{k}$ perpendicular to a infinite slab of scattering medium of thickness $L.$
Since the experiment is performed on an almost perfect two-level atomic ensemble, the vectorial nature of the fields -- i.e. the light polarization -- is irrelevant. Note that we do not attempt here to describe the field (multiply) scattered by the medium in other modes; the linearity of the equations implies that it will not affect the coherent outgoing beam.
In the stationary regime where the incoming laser is at fixed frequency $\omega,$ the outgoing electric field of the laser is proportional to the incoming field:
\begin{equation}
 \tEout(\omega) = \tEin(\omega) \exp\left(i\frac{n(\omega) \omega  L}{c}\right)
\label{eq:trans}
\end{equation}
where $n(\omega)$ is the complex index of refraction and $c$ the light velocity. When the medium is diluted, the index of refraction
is related to the atomic density $\rho$ and the individual atomic polarizability $\alpha(\omega)$:
$n(\omega) = 1 + \rho \alpha(\omega)/2$ \cite{hecht1974optics}.
 The scattering mean free path $\ell(\omega)$  is directly related to the imaginary part of the polarizability:
$\ell=c/[\rho\omega\Im(\alpha(\omega))],$ and the intensity transmission is simply $e^{-b},$
with $b=L/\ell(\omega)$ the optical thickness of the medium.

From now on, we consider the case of resonant scatterers -- typically isolated atomic scatterers -- which present
a narrow resonance centered around $\omega_0$ with a width $\Gamma\ll\omega_0,$ such that the atomic polarizability
is proportional to  $1/(\omega-\omega_0+i\Gamma/2)$ \cite{loudon2000quantum}. If the laser beam is switched off much faster than  $\Gamma^{-1},$ the atomic dipoles can be considered
as frozen during the switching, implying that they will continue to radiate the same field immediately
after the switching off, slowing decaying over a typical time scale $\Gamma^{-1},$ thus creating the coherent flash of light.
The preceding expression for the atomic polarizability is valid for atoms at rest. For moving atoms,
it is described by the convolution of the atomic polarizability shifted
by Doppler effect $1/(\omega-kv-\omega_0+i\Gamma/2)$ (with $v$ the atomic velocity along the laser axis) with the velocity distribution.

\begin{figure}
\begin{center}
\includegraphics[scale=0.70]{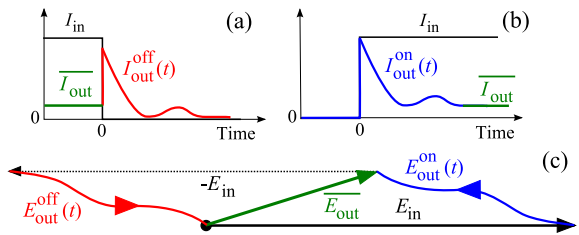}
\caption{Schematic view of the coherent transmission. (a) $\&$ (b): Two complementary "off" and "on" experiments, where
a monochromatic laser is abruptly switched off or on.  (c) Schematic temporal evolution of the outgoing electric field in the complex plane, for the two complementary experiments. In this example, the electric field
$\Eout$ is not in phase with the ingoing field $\Ein$. For a resonant laser, the theoretical prediction,
essentially confirmed by the experimental observations, see fig.~\ref{Phase}, is that the phase vanishes.}
\label{Complementarite}
\end{center}
\end{figure}

In order to analyze the transient phenomena, we make a Fourier
decomposition of the incoming and outgoing electric fields. In particular we consider two complementary experiments: in the first one (``off'' experiment), a resonant laser beam is sent on the medium during a long time, then abruptly switched off, while in the ``on'' experiment, the same beam is simply switched on abruptly and remains on forever~\cite{ducloy1977compl} (see a schematic description in fig.~\ref{Complementarite}).
The corresponding Fourier components of the incoming fields are simply:
\begin{equation}
\tEinoff(\omega) = \frac{iE_0/2\pi}{\omega-\omega_0+i0^+}\ \ \ \ \tEinon(\omega) = \frac{iE_0/2\pi}{\omega-\omega_0-i0^+}
\label{eq:ein}
\end{equation}

The temporal dependence of the outgoing electric field is readily obtained by injecting eqs.~(\ref{eq:ein}) in the
stationary response, eq.~(\ref{eq:trans}), and Fourier transforming back from frequency to time. The superposition of the ``on"
and ``off" incoming fields is a monochromatic field at frequency $\omega_0,$ which produces a
stationary coherent outgoing field $\bEout$ at the same frequency.
The coherently transmitted fields in the ``on'' and ``off'' experiments are thus such that:
\begin{equation}
 \Eouton(t)+\Eoutoff(t)=\bEout
\label{eq:complementarite}
\end{equation}

When the laser is switched on abruptly, it takes some time for the atomic dipoles to build up and radiate
a coherent field antagonist to the incoming one, meaning that the medium is initially transparent, reaching its stationary transmission over a time of the order of $\Gamma^{-1}.$
This phenomenon -- known as optical nutation \cite{brewer1971photo} -- is also schematically described in fig.~\ref{Complementarite}.
For large optical thickness, $\bEout$ is negligibly small compared
to the incoming field, so that $\Eoutoff(t=0^+)=\bEout - \Eouton(t=0^+) \approx -\Ein$:  the coherent flash of light outgoing from the medium is then as intense as the incoming laser (with a $\pi$ relative phase), although the latter is already switched off!

The detailed calculation of these various effects is possible, and will be presented in a forthcoming publication \cite{romain_paper_long}. The imaginary part of the atomic polarizability is unchanged when the detuning changes sign while its real part
changes sign. This implies that $\tEout(\omega_0+\delta)\approx \tEout^*(\omega_0-\delta)$
and that, for a laser frequency resonant with the atomic frequency, both $\Eouton(t)$ and $\Eoutoff(t)$
are real quantities, in phase (or in opposition) with the incoming field $\Ein,$ as experimentally observed,
see below.
The shape of the coherent flash of light depends on the various parameters, but it turns out
that, at resonance, the initial decay rate (at time $t=0^+$) has a simple expression:
\begin{equation}
\tau =  \frac{\Ioutoff(t=0^+)}{-\frac{d\Ioutoff}{dt}(t=0^+)} = \frac{1-\exp(-b/2)}{b/2}\ g\left(\frac{k\vbar}{\Gamma}\right)\ \Gamma^{-1}
\label{decay_time}
\end{equation}
where $g$ is related to the complementary error function \cite{abramowitz1964handbook} (assuming a Boltzman
velocity distribution):
\begin{equation}
 g(x) = \sqrt{\frac{\pi}{8}} \frac{1}{x} \exp\left(\frac{1}{8x^2}\right) \erfc\left(\frac{1}{\sqrt{8}x}\right)
\label{eq:defg}
\end{equation}
The first term in eq.~(\ref{decay_time}) expresses the shortening of the coherent flash in optically thick media,
while the $g$ function (always smaller than or equal to unity) the effect of the atomic velocity.
For large atomic velocity, $g(x)\propto 1/x,$ so that the decay time
scales as $1/k\vbar,$ as expected from the width of the atomic absorption profile.

\begin{figure}
\begin{center}
\includegraphics[scale=0.6]{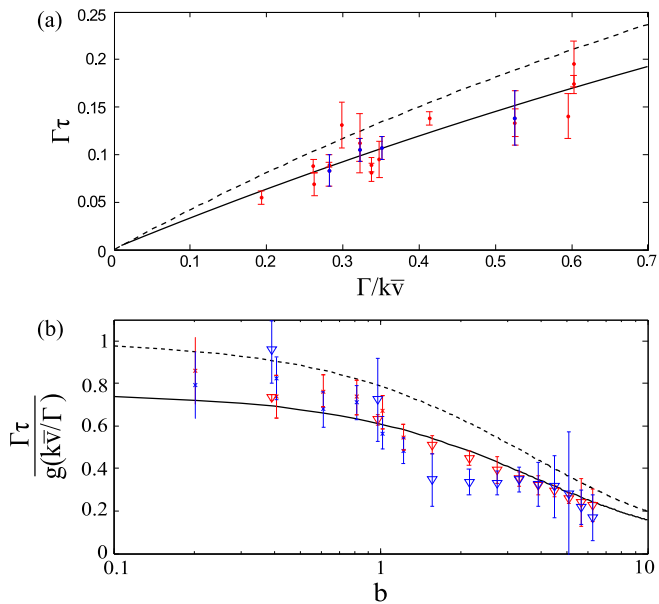}
\caption{(color online) Characteristic decay time of the coherent transmission at the laser ignition (blue symbols) and extinction (red symbols). The dashed line is the analytic prediction, eq.~(\ref{decay_time}), based on the short time
decay of the coherent pulse, while the
solid curve corresponds to a fit of the full theoretical curve with a decreasing
exponential. (a): Data taken as a function of the average
atomic velocity, at approximately constant optical thickness $b=1.2-1.6.$ Theoretical curves are computed for $b=1.5.$ (b): Data taken as a function of the optical thickness $b$ at two temperatures: $T=1.0(2)\,\upmu$K or $k\vbar=1.8\Gamma$ (stars) and $T=3.8(4)\,\upmu$K or $k\vbar=3.7\Gamma$ (triangles).
}
\label{Carac_time}
\end{center}
\end{figure}

These predictions are confronted to the experimental observations in fig.~\ref{Carac_time}: the agreement
is good, both when the atomic velocity or the optical thickness are varied.
Note that the characteristic decay time of the experimentally observed coherent flash is measured
by fitting it with a decreasing exponential, although the shape decreases faster than an exponential.
This is why expression (\ref{decay_time}) slightly overestimates the decay time. When the same fitting
procedure is used on the theoretical curves, the characteristic time is typically $20\%$ smaller
and agrees perfectly well with the experimental observations.

\begin{figure}
\begin{center}
\includegraphics[scale=0.95]{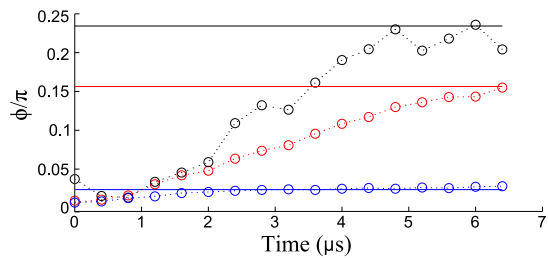}
\caption{Evolution of the relative phase $\phi$ between $\Eouton(t)$ and $\Ein$ as function of time for $b=0.3$ (blue circles), $b=1.2$ (red circles) and $b=3.5$ (black circles). The horizontal lines are the stationary values.}
\label{Phase}
\end{center}
\end{figure}

The experimental decay times, shown in fig.~\ref{Carac_time}, are extracted
assuming a relative phase $\phi=0$ between $\Eouton(t)$ and $\Ein$ as suggested by the theoretical calculation. However, using the complementarity of the "on" and "off" experiments, one can extract $\phi(t)$ using the expression (\ref{eq:complementarite}), from the measurement of the intensity in both the "off" and the "on" experiment. We have checked, see fig.~\ref{Phase},
that $\phi(t)$ remains small at short time ($t<2\,\upmu\textrm{s}$), up to the typical decay time. At longer time, the phase shift is more important especially at large optical thickness. This phase shift, not predicted by the theory developed here, could be due to some experimental imperfections such as a systematic frequency offset of the probe laser with respect to the atomic resonance, or to some inherent limitations of the experiment like mechanical effect when the probe laser is on. It could also come from a
breakdown of the independent scattering assumption since the medium is not extremely dilute ($\rho\lambda^3\simeq 0.1$). This interesting point is left
for future investigations.

In the actual experiment, the atomic medium is not
an infinite slab, but a spherical cloud with a Gaussian shaped atomic density. When we look
at the coherent beam, only forward propagation is important so that the relevant parameter is the optical thickness
along an optical ray. It is maximum when the ray passes through the center of the atomic cloud and decays like
$b(r)=b_{\mathrm{max}} \exp(-r^2/2R^2)$ as a function of the transverse distance $r$ from the center
($R$ being the radius of the cloud). The coherent flash is thus expected to decay faster at the cloud center
-- the larger the optical thickness, the shorter the decay time -- than in the external layers.
This is visible in the experimental records, producing a spatio-temporal dynamics with a characteristic ring shape visible in fig.~\ref{Time_Evo}(a)
at time $t=2.4\,\upmu\textrm{s}$. At large optical thickness, the broadening and the distortion -- with respect to the natural Lorentzian shape -- of the spectral absorption window [see eq.~(\ref{eq:trans})] leads to an oscillatory temporal evolution of the flash, clearly observed in the experiment, see fig.~\ref{Time_Evo}(b), with a minimum of the intensity in the center at $t=2.4\,\upmu\textrm{s}$, followed by a revival at $t=4.8\,\upmu\textrm{s}$. This reenforces the ring shape.
We have also performed \textit{ab initio} calculations of the electromagnetic field coherently
transmitted through a set of randomly placed atomic dipoles~\cite{LAX-1952} --
using the full geometry of the atomic cloud -- which confirm this interpretation.

To summarize, thanks to the slow response time of the Strontium intercombination line,
we have studied the spatio-temporal dynamics of the coherent transmission of a resonant laser beam
across a scattering medium. We have shown the existence of a strong coherent flash of light
following the extinction of the laser, and measured its characteristic properties vs. temperature and optical thickness of the medium. Using the complementary information of the laser ignition, we have reconstructed the phase of the forward transmitted electric field.

M. Ducloy and C. Miniatura are acknowledged for fruitful discussions.


\end{document}